# Calibrating the self-thinning frontier

Jerome K. Vanclay[1] and Peter J. Sands[2]
[1] Southern Cross University, PO Box 157, Lismore NSW 2480, Australia
[2] 39 Oakleigh Av, Taroona, Tasmania, Australia 7053

**Correspondence:** JVanclay@scu.edu.au, Tel: +61 2 6620 3147, Fax: +61 2 6621 2669

**Abstract**
Calibration of the self-thinning frontier in even-aged monocultures is hampered by scarce data and by subjective decisions about the proximity of data to the frontier. We present a simple model that applies to observations of the full trajectory of stand mean diameter across a range of densities not close to the frontier. Development of the model is based on a consideration of the slope $s = \ln(N_t/N_{t-1})/\ln(D_t/D_{t-1})$ of a log-transformed plot of stocking $N_t$ and mean stem diameter $D_t$ at time $t$. This avoids the need for subjective decisions about limiting density and allows the use of abundant data further from the self-thinning frontier. The model can be solved analytically and yields equations for the stocking and the stand basal area as an explicit function of stem diameter. It predicts that self-thinning may be regulated by the maximum basal area with a slope of -2. The significance of other predictor variables offers an effective test of competing self-thinning theories such Yoda's -3/2 power rule and Reineke's stand density index.

1. **Introduction**
The theory of limiting density (Reineke, 1933) and self-thinning (Yoda *et al.,* 1963) in even-aged monocultures continues to attract attention (Pretzsch, 2002; Bi, 2004; Pretzsch and Biber, 2005; Reynolds and Ford, 2005) decades after being proposed, but an efficient and satisfactory procedure to calibrate the self-thinning frontier remains elusive (Zhang *et al.,* 2005; Vanderschaaf and Burkhart, 2007). Many methods are hampered by the need to make a subjective selection of samples considered to be representative and at or near the frontier (Zhang et al. 2005). Despite doubts about the validity of the concept (Reynolds and Ford, 2005), there remains a need to reduce this subjectivity because the concept is widely applied in forest research and management.

A key principle implicit the Reineke and Yoda propositions is that any arrangement of regular objects in a single layer within a confined area has a volume-area relationship in which the number $N$ of objects and their volume $V$ exhibit a power curve $V \propto N^{-3/2}$ (Pretzsch, 2002) – or equivalently, that the relationship between size $S$ and number is $N \propto S^{-2}$ (e.g., where S is the radius of identical spheres arranged on a plane surface). In a frequently cited paper, Yoda *et al.* (1963) observed that this Euclidean fundamental applies to herbaceous plants. Decades earlier, Reineke (1933) observed a slope of -1.605 in the size-stocking power curve for several north American conifers, an observation at odds with the -2 slope indicated by Yoda's proposition. Within a few years, MacKinney and Chaiken (1935) completed a statistical analysis of Reineke's original data and estimated the slope as -1.707. More recently, Pretzsch and Biber (2005) have argued that the slope is species-specific. West *et al.* (1997) have advocated a slope of -4/3, but their analysis has been challenged (Kozlowski and Konarzewski, 2004; Stegen and White, 2008). Many subsequent studies have examined whether these trends do, or do not exist in plant communities (for recent reviews, see e.g., Reynolds and Ford, 2005; Shaw, 2006).

Several characteristics of the self-thinning frontier hamper empirical study and calibration. The frontier, rather like a black hole, is not visible directly, but must be inferred indirectly from the death of individuals as a stand approaches the frontier. The self-thinning frontier is not a constant unyielding barrier, but is more like a water table that fluctuates with the seasons, manifesting itself differently at times according to limiting resources. As a result, the frontier can be estimated only indirectly, approximately, and asymptotically.

Further complications arise from the empirical relationships that are used to describe the frontier. Some discrepancies may arise because the space occupied by a tree is determined in part by its crown, rather than by the stem diameter used as the basis for Reineke's stand density index. If the relationship between stem diameter and crown diameter is $C=\beta D^{0.8}$ (in the case of Reineke's estimate), then there is no conflict, and the stand density index complies with the expected Euclidean trend and with the crown competition factor (Krajicek *et al.*, 1961). Smith and Hann (1984) observed that when there is an allometric relationship between diameter and volume, $V=\beta D^{2.4}$, Reineke's and Yoda's hypotheses concur. Recently, Zeide (2005) has suggested a modification to Reineke's equation to better account for tree size and packing, and Garcia (2009) has advocated the merits of an analogous approach based on top height rather than diameter.

It can be demonstrated empirically that the slope of the number-size power curve is unaffected by packing (i.e., regular versus random placement of trees), and by any lag that may occur while neighbours grow into a space created by the death of a plant. Any departure from the nominal slope of -2 is primarily due to the allometric relationship between stem diameter and crown size, or more specifically, between stem diameter and the space needed to satisfy photosynthetic and respiratory demands. Notwithstanding claims by Enquist and Niklas (2001), it is reasonable to expect that trees in different environments may exhibit different size:space relationships (Morris, 2002), influenced by the space needed to capture limiting resources.

Yoda's self-thinning line and Reineke's stand density index are useful and widely used in plantation growth models to predict natural mortality (e.g., Monserud *et al.*, 2005), including in process-based models (e.g., Landsberg and Waring 1997). Calibrating these relationships is notoriously difficult and demanding of data, and this paper considers an alternative approach to estimate self-thinning trends such as Yoda's and Reineke's lines. Rather than selecting data believed to be at the self-thinning frontier, it is expedient to examine the full trajectory of stand mean diameter across a range of densities by examining $s = \ln(N_t/N_{t-1})/\ln(D_t/D_{t-1})$, where $N_t$ and $D_t$ are the stocking and mean diameter at time $t$. We present a simple model based on the assumption, supported by observations on many stands, that $s$ can be approximated by a power function of the current stand basal area. The resulting model can be solved analytically to give explicit equations for both stocking and basal area as a function of diameter. The model has two parameters: the maximum basal area attained during self-thinning, and the power, which determines how rapidly a stand approaches the self-thinning line. The model is very easy to fit to observed stocking v. diameter data, and its use avoids the need for subjective decisions about limiting density and allows the use of abundant data further from the frontier.

**2. Materials and Methods**
The assumption usually made in interpreting and applying the self-thinning line is that growth slows and mortality increases as a forest stand approaches the limiting stand density, but this assumption is rarely taken into account explicitly when estimating the frontier. The self-thinning frontier is usually estimated by subjectively selecting data considered to be close to the frontier, but an alternative is to examine the first differences of successive observations of forest condition. Others (e.g., Roderick and Barnes, 2004; Pretzsch and Biber, 2005; Zhang *et al.*, 2005; Vanderschaaf and Burkhart, 2007) have examined first-differences, but have not commented on the evolution of these trajectories as they approach the frontier.

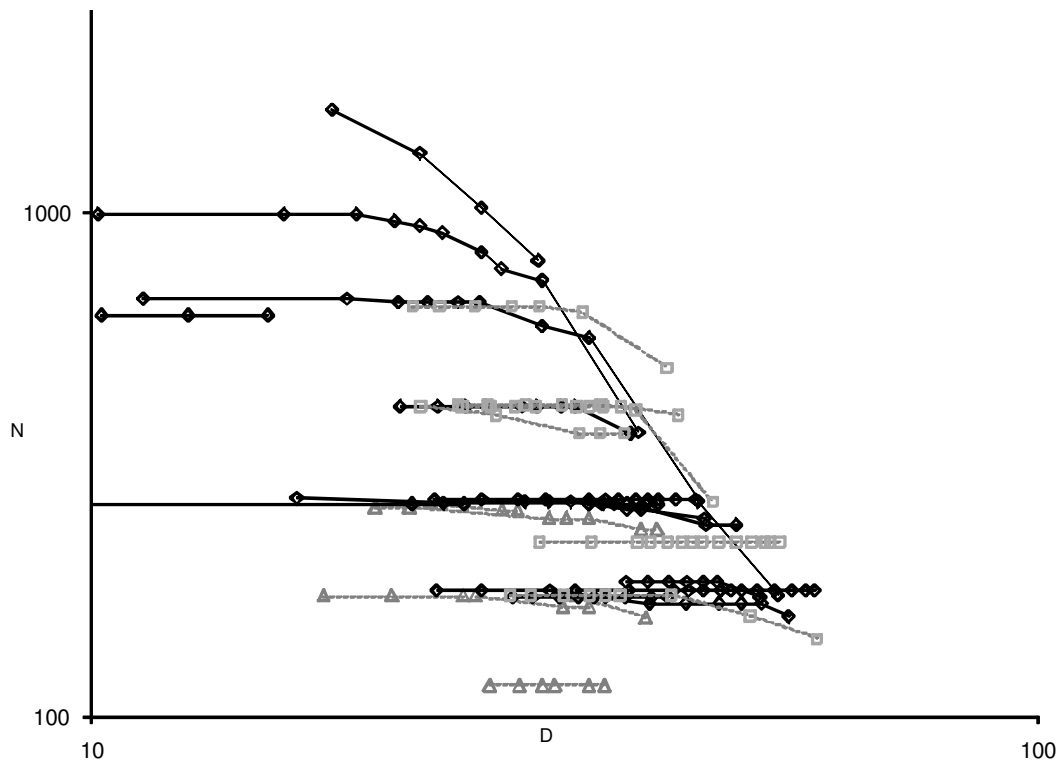

**Figure 1**. Self-thinning trends in *Eucalyptus pilularis* forests in Queensland, illustrated as a log-log garph. High productivity plots marked with squares (□), typical plots marked with diamonds (◊), and low productivity plots marked with triangles (Δ). N in stems/ha, and D in mm dbh.

The slope *s* of the trajectory observed on the log-log graph illustrated in Fig. 1 can be estimated as the first difference of successive observations

$$s = \frac{d(\ln N)}{d(\ln D)} \approx \frac{\ln(N_t / N_{t-1})}{\ln(D_t / D_{t-1})},\qquad(1)$$

where $N_t$ is the number of individuals and $D_t$ is their mean size (diameter at breast height, 1.3 m) at time *t*. This formulation expresses the slope in the form considered by Reineke (viz. $N=f(D)$), the inverse of the form considered by Yoda ($V=f(N)$). Note that *s* is not a constant, but defines a trajectory, and is expected to have a near-zero value in stands with low densities, and to increase and approach a limiting slope *s\** as density increases. According to the Reineke and Yoda propositions, *s\** may be in the range -1.6 to -2. It is useful to examine full trajectory of *s* across a wide range of densities because data are often more abundant further from the self-thinning frontier, and this avoids the need for subjective decisions about proximity to the frontier. In many cases, this approach is more faithful to the available data, which often informs how forest stands approach the self-thinning frontier, rather than how they behave at the frontier itself.

**Table 1**. Characteristics of 29 plots of *Eucalyptus pilularis* used to examine the self-thinning response.

| Attribute | Minimum | Mean | Maximum |
|---|---|---|---|
| Establishment date | 1923 | 1928 | 1971 |
| Stand age (years) | 1 | 35 | 63 |
| No of measures | 8 | 19 | 31 |
| Site productivity | 22 | 32 | 41 |
| Stem diameter (cm) | 1 | 32 | 80 |
| Basal area (m$^2$/ha) | 1 | 31 | 71 |
| Stems/ha | 83 | 365 | 1594 |

The utility of this approach was examined using data from several sources, but is illustrated primarily with *Eucalyptus pilularis* Sm. data (Table 1) from a national collection of growth and yield data from eight eucalypt species growing in even-aged, monoculture forest (West and Mattay, 1993; Mattay and West, 1994). Plots that had not been re- measured, and intervals involving harvesting or artificial thinning were omitted from the analysis. Measurement intervals in these data varied greatly (3 months to 14 years), so intervals were combined to create intervals $\geq 2$ years with $D_{t+1} - D_t \geq 1$ cm to avoid the high variance in estimates of *s* that may arise with pairs of observations with minimal increment.

Death in trees may not be conspicuous and sudden. Assessors may regard a tree as 'dead', only to discover green shoots at the next measure, before death is finally confirmed at some subsequent remeasure. In addition, death is often clustered in time and space (Vanclay, 1991a), so data derived from short intervals may exhibit a stepped approach to the self-thinning frontier. Thus, in dense stands (>30 m$^2$/ha), the few intervals that did not include mortality were combined to create intervals with $N_{t+1} < N_t$. Figure 1 illustrates the resulting data for published *Eucalyptus pilularis* in Queensland (Mattay and West, 1994). In Figure 1, it is evident that the self-thinning frontier may depend on site productivity estimated from predominant height at age 35 years (Skovsgaard and Vanclay, 2008). Several researchers (e.g., Bi, 2001; Larsen et al., 2008; Wieskittel et al., 2009) have previously observed that site productivity influences the self-thinning frontier.

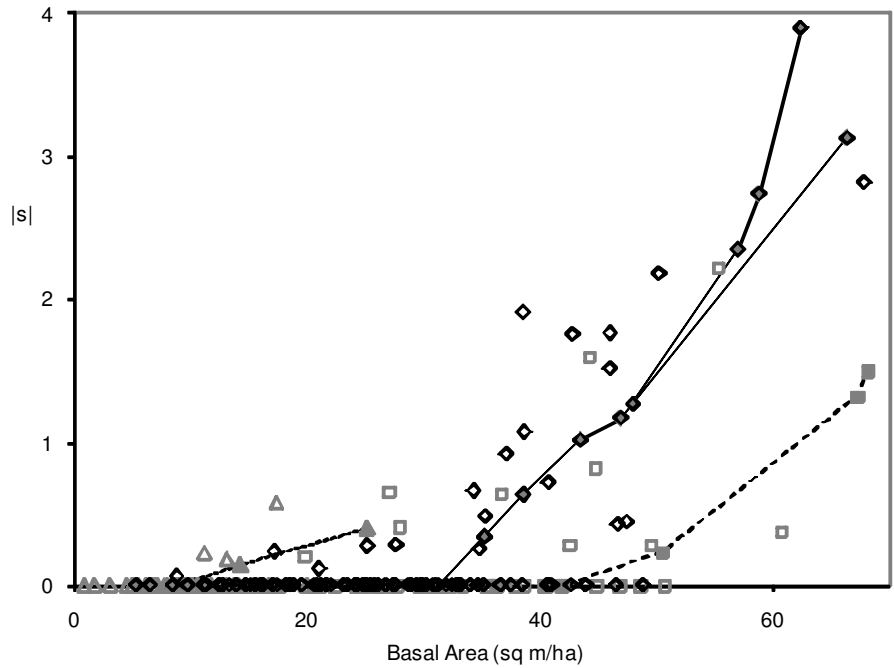

**Figure 2**. Absolute value of the slope of the self-thinning trend s = ln($N_2/N_1$)/ln($D_2/D_1$) plotted against stand basal area for *Eucalyptus pilularis* in Queensland, with four lines illustrating the self-thinning trajectories of four plots of low (▲), average (♦) and high (■) site quality

Figure 2 shows the values of *s* derived from this set of data, along with the actual trajectory for *s* obtained from four specific stands. The large number of zero values for *s* arise in part from stands that were not (yet) self-thinning, and in part from the inherent random nature of death. However, the specific trajectories show a strong correlation with stand basal area, and are well represented by a simple power function of basal area. We show below that this offers a way to predict self-thinning trajectories for stands which do not have a long history of repeated measurement. Other work (e.g., Vanclay, 1991b) suggests that stand basal area should provide a good basis for predicting *s*, but other candidates could include leaf area index (Hamilton *et al.*, 1995; Innes *et al.*, 2005), aggregate height (Fei *et al.*, 2006), or top height (Garcia, 2009). The possibility that *s* may be estimated adequately from basal area alone implies that the self-thinning frontier will have a slope $s^* = -2$, but the inclusion of additional predictor variables such as ln(*D*) are needed to provide $s^* > -2$ consistent with Yoda's and Reineke's propositions. If we assume that *s* can be approximated by a power function of basal area $G = \pi N(D/200)^2$ alone, i.e.

$$s = -2(G/G_x)^n \qquad (2)$$

then equation (1) can be integrated (see Appendix) to give explicit equations for stem number and stand basal area as explicit functions of current stem diameter. In equation (2), *n* is a power and $G_x$ is the basal area at which $s = -2$ (and also the maximum basal predicted by the model). The integrated equations of the model are

$$N(D) = N_0 \Big/ \left[1+(n_0 D^2 / G_x)^n\right]^{\frac{1}{n}}$$
$$G(D) = G_x \Big/ \left[1+(G_x / n_0 D^2)^n\right]^{\frac{1}{n}} \tag{3}$$

where $N_0$ is initial stocking (i.e. stocking for small $D$) and $n_0 = \pi N_0/40000$. For large $D$, i.e. when self-thinning is occurring, these equations give $N = (40000/\pi)(G_x/D^2)$, i.e. $N \propto D^{-2}$, and $G = G_x$.

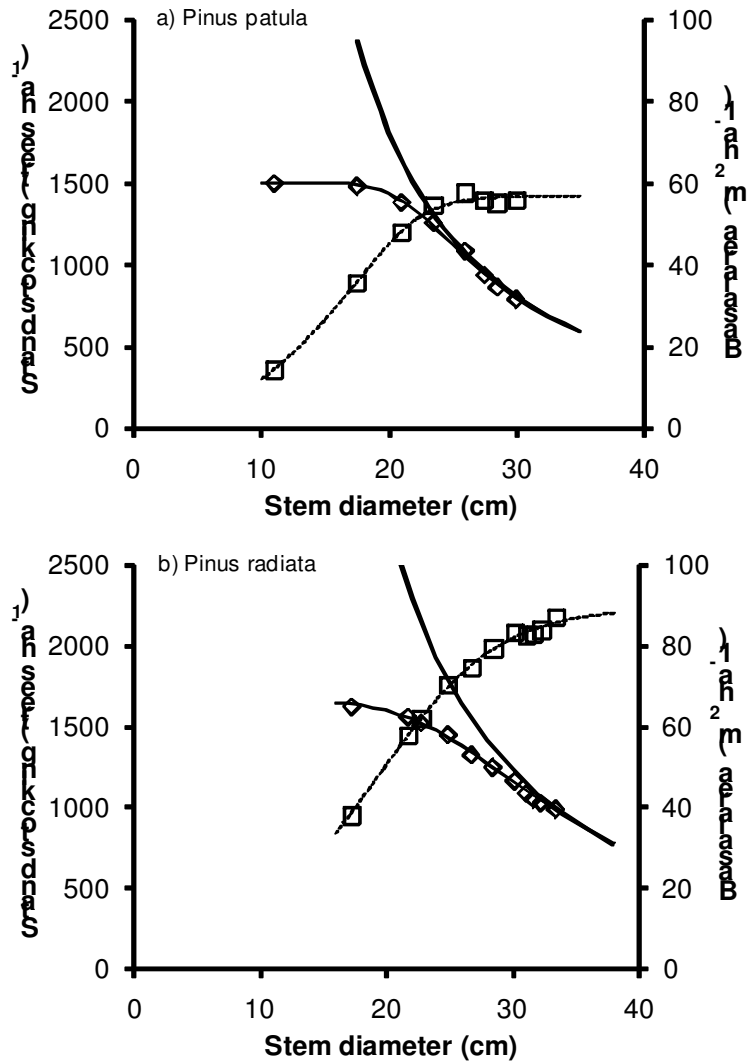

**Figure 3.** Results of applying equations (3) to self thinning trajectories of two stands. (a) *Pinus patula* grown in South Africa (Dye, 2001), with $n = 6$ and $G_x = 57$. (b) *Pinus radiata* grown in South Australia (this study), with $n = 3.5$ and $G_x = 90$. The data were fitted by eye by setting $G_x$ and then varying $n$. The bold line represents an estimate of the self-thinning frontier; diamonds indicate the actual stand size-density trajectory (left axis) and squares indicate the basal area development (right axis).

## 3. Results

The ability of the model given by equations (3) to fit individual self-thinning trajectories is illustrated in Figure 3 (showing the un-transformed data) for two distinct stands of different species grown in two different locations. The parameters $n$ and $G_x$ were estimated by fitting the model to the data by eye. The fits are fairly insensitive to the power $n$ and in the following analysis we apply the same power ($n = 3$) to a large number of stands, although this value was not estimated in a rigorous manner.

The case of the *Eucalyptus pilularis* data shown in Figure 2 can be modelled with the simple equation $s = -0.436(G/H)^3$, where $H$ is the expected height of predominant trees at age 35 years. Although simple, this is an adequate model, with a small standard error (0.021, P<0.001), and no evidence of lack of fit (P=0.35; Weisberg, 2005). Other predictor variables such as $\ln(D)$ were not significant (P=0.2), suggesting that self-thinning in this species is correlated with stand basal area, and that maximum basal area (Assmann, 1970; Sterba and Monserud, 1993; Skovsgaard and Vanclay, 2008) is a sufficient concept to explain self-thinning and offering no support for Reineke's and Yoda's propositions. Figure 4 illustrates the self-thinning trend implied by this simple equation, and confirms the adequate fit to the data.

This ability to make reasonable predictions of self-thinning by predicting the slope $s$ from a power of basal area was confirmed with other published (Mattay and West, 1994) and unpublished data. Figure 5 illustrates estimates of self-thinning in *Pinus radiata* D.Don in South Australia obtained for $n = 3$ and $G_x = 95$ (based on 4 plots aged 11-62 with site quality III-IV). This suggests that the first difference approach as implemented using equations (3) is an efficient way to estimate self-thinning in crowded stands, and that in many cases, basal area is a useful predictor of the trajectory.

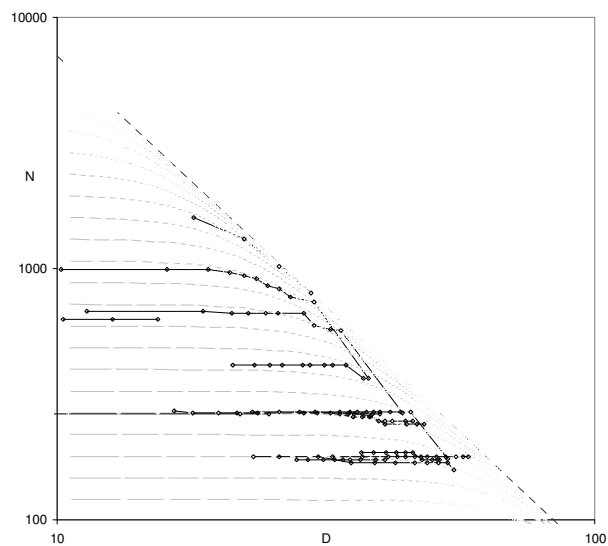

**Figure 4**. Self-thinning trends in *Eucalyptus pilularis* of near-average site productivity (28-38 m predominant height at age 35 years). Black lines are observed data (Mattay and West, 1994). Curved horizontal lines are constructed from estimates of $s$ with $n = 3$ and $G_x = 1.79H$ where $H$ is the expected height of predominant trees at age 35 years. Each grey dash represents the estimated 2 year increment. Diagonal dashed line represents a stand basal area of 55 $m^2$/ha.

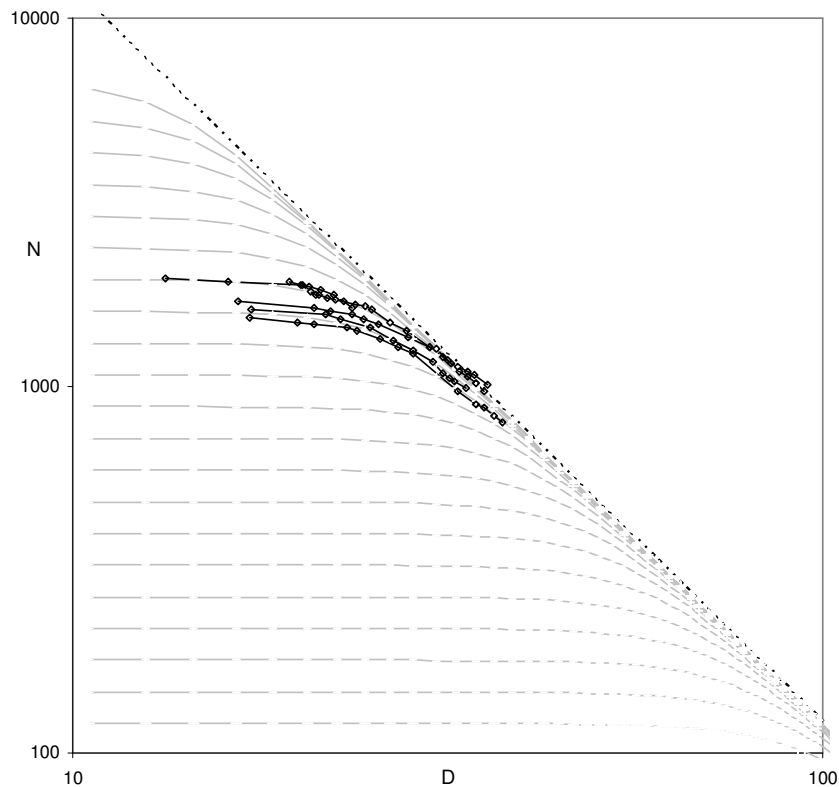

**Figure 5**. Predicted and observed self-thinning of *Pinus radiata* in South Australia spanning an age range 11-62 years and site quality III-IV. Dotted diagonal line is $G = 95$ m$^2$/ha, and $s$ is estimated with $n = 3$ and $G_x = 95$.

Estimating the trajectory solely from basal area leads to a series of self thinning lines (Figures 4 and 5) that converge toward a site-dependent maximum stand basal area with $s^* = -2$. It is appropriate to examine other predictor variables such as $s = \beta_0 + \beta_1 G^n + \beta_2 \ln D$, which could accommodate $s^* \neq -2$ and (depending on the value of $\beta_2$) support the Reineke-MacKinney proposition that $s^* > -2$. This approach offers an efficient and non-subjective way to estimate the slope $s$, and to test the adequacy of Reineke's and Yoda's propositions.

## 4. Discussion

A slope of $s^* = -2$ is a direct consequence of assuming $s$ is a power function of basal area, with no other explanatory variables, and implies that self-thinning is regulated by maximum stand basal area rather than according to Reineke's proposition. A self-thinning frontier with a slope other than $s^* = -2$ (e.g., -3/2 as proposed by Yoda) implies that other variables additional to basal area are required to predict $s^*$.

A slope of $s^* = -2$ is consistent with Yoda's proposition if $V = \beta D^3$, which may apply to some small organisms but which rarely applies to forest trees. More generally, the slope $s^* = -2$ imposes the constraint $n_S = -2n_t$ where $n_S$ is the allometric power for stem volume or mass as a function of diameter, and $n_t$ is the slope of the log-transformed stem mass v. stand density self thinning line. If $n_t = -3/2$, as often assumed, then $n_S = 3$. The Reineke-MacKinney proposition holds only if basal area is an inadequate estimator of $s$ and requires $\ln D$ as an additional predictor variable. Reineke's proposition arises if $s$ is calibrated as $s = \beta_0 + \beta_1 G^n - 0.4\ln D$, but this value was not evident in the data examined.

The analysis of the full trajectory using our model appears to be a practical and efficient way to estimate the self-thinning frontier. It minimizes the need for subjective decisions, and allows efficient statistical testing of Yoda's and Reineke's propositions. This approach suggests that in many cases, the concept of maximum stand basal area may be a more practical and parsimonious explanation of mortality in even-aged forest monocultures.

## Acknowledgements

Adrian Goodwin of Bushlogic offered constructive comment on a draft of this paper.

## Appendix: Derivation of Equations (3)

Assume that the slope $s$ of the $\ln N$ v. $\ln D$ curve is a power function of basal area $G$:

$$s = \frac{d(\ln N)}{d(\ln D)} = \frac{D}{N}\frac{dN}{dD} = -2(G/G_x)^n, \tag{4}$$

where $n$ is a power and $G_x$ is the basal area at which the slope $s$ is 2. The units are assumed to be $N$ in trees ha$^{-1}$, $D$ in cm and $G$ in m$^2$ ha$^{-1}$, so

$$G = \pi N \left(\frac{D}{200}\right)^2. \tag{5}$$

Substitute (5) into (4) to get

$$\frac{D}{N}\frac{dN}{dD} = -\gamma N^n D^{2n}, \tag{6}$$

where $\gamma = 2(\pi/40000 G_x)^n$. Rearrange (6) so that $N$ is on the left and $D$ on the right to give an equation that can be directly integrated using the rule that the integral of $x^{n-1}$ is $x^n/n$. Integration gives

$$\frac{1}{n}N = C + \frac{\gamma}{2n}D^{2n}, \tag{7}$$

where $C$ is the constant of integration which is set using the initial stem number, assumed to be $N_0$ when $D = 0$. The final result for stocking as a function of DBH is

$$N(D) = N_0 \left/ \left[1 + (n_0 D^2/G_x)^n\right]^{\frac{1}{n}} \right. . \tag{8}$$

where $n_0 = \pi N_0/40000$. The corresponding basal area follows by combining (5) and (8):

$$G(D) = \pi N \left(\frac{D}{200}\right)^2 = \frac{\pi}{40000}\left(\frac{(N_0 D^2)^n}{1+(n_0 D^2/G_x)^n}\right)^{\frac{1}{n}} \tag{9}$$

$$= G_x \left/ \left[1 + (G_x/n_0 D^2)^n\right]^{\frac{1}{n}} \right. .$$